\pdfoutput=1
\documentclass[galaxies,article,referee,oneauthor,pdftex,12pt,a4paper]{mdpi}
\usepackage{graphicx}
\usepackage{url}
\makeatletter
\renewcommand\@biblabel[1]{#1. }
\makeatother
\def\sun{\odot}
\def\h0units{\mathrm{km\,s^{-1}\,Mpc^{-1}}}
\def\cunits{\mathrm{km\,s^{-1}}}
\newcommand{\om}{\Omega_{\rm M}}

\newcommand{\dl}{d_{\rm{L}}}

\def\aap{A\&A\,  }%% Astronomy and Astrophysics
\def\aj{AJ  }%% The Astronomical Journal
\def\apj{ApJ\,  }%% Astrophysical Journal
\def\apjs{ApJS  }%% Astrophysical Journal, Supplement
\def\apss{Astrophysics and Space Science  }%Astrophysics and Space Science
\def\mnras{MNRAS\,  }%% Monthly Notices of the RAS

\lastpage{x}
\doinum{10.3390/------}
\pubvolume{xx}
\pubyear{2015}
\history{Received: xx / Accepted: xx / Published: xx}
\Title
{
A left and right  truncated Schechter
luminosity function for quasars 
}

\Author{Lorenzo Zaninetti $^{1,}$}

\address
{
$^{1}$
Physics Department,
 via P.Giuria 1,\\ I-10125 Turin,Italy 
}

\corres
{
zaninetti@ph.unito.it
}

\abstract
{
The  luminosity function for quasars (QSOs) is usually 
fitted by a Schechter  function.
The dependence of the number of quasars 
on the redshift, both in the low and high luminosity
regions, requires the  inclusion of a lower and upper 
boundary in  the Schechter function.
The normalization of the truncated 
Schechter function is forced to be the same as that for the Schechter function,
and an analytical form for the average value is derived.
Three astrophysical applications for QSOs are provided:
deduction of the parameters at low redshifts,
behavior of the average absolute magnitude at high 
redshifts, and the location (in redshift) 
of  the photometric maximum 
as a function of the selected apparent magnitude.
The  truncated Schechter function
with the double power law and an 
improved Schechter function are compared as luminosity functions for QSOs.
The chosen cosmological framework is that of the flat cosmology,
for which we provided the luminosity distance, the inverse 
relation for the luminosity distance, and the distance modulus.
}

\begin{document}

\section{Introduction}

The Schechter function 
was first introduced in order to model 
the luminosity function (LF)  
for  galaxies, see
\cite{schechter},
and later was used to model 
the LF for quasars (QSOs), 
see \cite{Warren1994,Goldschmidt1998}.
Over the years, 
other LFs for galaxies  have been suggested, such as a 
two-component Schechter-like LF, 
see~\cite{Driver1996},
the hybrid Schechter+power-law LF
to fit the faint end of the K-band, see~\cite{Bell2003},
and the  double  Schechter LF, 
see~\citet{Blanton_2005}. 
In order  to  improve the flexibility
at the bright  end, a new parameter
$\eta$ was  introduced in the Schechter LF,
see \cite{Alcaniz2004}.
The above discussion suggests the introduction 
of finite boundaries for the Schechter LF 
rather than the usual zero and infinity.
As a practical example the 
most luminous
QSOs have  absolute magnitude  
$M_{b_j} \approx -28$ or the luminosity is not $\infty$  
and 
the less luminous
QSOs have have absolute magnitude  
$M_{b_j} \approx -20$
or the luminosity is not zero, see Figure 19 in 
\cite{Croom2004} .
A physical source of truncation at 
the low luminosity boundary ( high absolute magnitude ) 
is the fact that with increasing redshift
the less luminous QSOs progressively 
disappear. In other words the upper boundary 
in absolute magnitude for QSOs  is function of the redshift.

The  suggestion to introduce two 
boundaries in a probability density function (PDF)    
is not new 
and, as an example, 
\cite{Coffey2000} considered a doubly-truncated gamma PDF
restricted by both a
lower (l) and upper (u) truncation.
A way to deduce a new truncated LF for galaxies or QSOs  
is to start from a truncated PDf and then to derive
the magnitude version.
This  approach has been used to deduce a left truncated 
beta LF, see \cite{Zaninetti2014d,Zaninetti2015a},
and a truncated gamma LF, see \cite{Zaninetti2016a}.

The main difference between LFs for galaxies and for QSOs 
is that in the first case, we have an LF for a unit volume 
of 1 $Mpc^3$ and in the second case we are speaking 
of an LF for unit volume but with a redshift dependence.
The dependence on the redshift complicates an analytical 
approach, because the  number of observed  QSOs 
at low luminosity decreases with the redshift and 
the highest observed luminosity increases 
with the redshift.
The first effect is connected with the Malmquist bias,
i.e. the average luminosity increases with
the redshift, and the  second one can be modeled by an empirical law.
The above  redshift dependence in the case of QSOs  
can be modeled by the double power law LF, see \cite{Boyle1988},
or by an improved   Schechter function, see \cite{Pei1995}. 
The present paper derives, in Section \ref{secflat}, 
the luminosity distance and 
the  distance modulus in a flat cosmology.
Section \ref{sectruncated} derives 
a truncated version of the 
Schechter LF. 
Section \ref{secqso}  applies the truncated Schechter LF
to QSOs, deriving the parameters  of the LF 
in the range of redshift $[0.3 , 0.5]$,
modeling the average absolute magnitude as a function 
of the redshift, and deriving the photometric maximum for a given
apparent magnitude as a function of the redshift.   

\section{The flat cosmology}

\label{secflat}
The {\it first} definition of the   
luminosity distance, $\dl$, in flat cosmology is
\begin{equation}
  \dl(z;c,H_0,\om) = \frac{c}{H_0} (1+z) \int_{\frac{1}{1+z}}^1
  \frac{da}{\sqrt{\om a + (1-\om) a^4}} \quad ,
  \label{lumdistflat}
\end{equation}
where $H_0$
is the Hubble constant expressed in     $\h0units$,
$c$ is the speed  of light expressed in $\cunits$,
$z$ is the redshift,
$a$ is the scale-factor,
and  $\om$ is
\begin{equation}
\om = \frac{8\pi\,G\,\rho_0}{3\,H_0^2}
\quad ,
\end{equation}
where $G$ is the Newtonian gravitational constant and
$\rho_0$ is the mass density at the present time,
see  eqn~(2.1) in \cite{Adachi2012}.
A {\it second} definition of the   
luminosity distance is 
\begin{equation}
  \dl(z;c,H_0,\om) = \frac{c}{H_0} (1+z) 
\int_{0}^{z}\!{\frac {1}{\sqrt { \left( 1+z \right) ^{3}{\it \om}+1
-{\it \om}}}}\,{\rm d}z
\quad ,
  \label{lumdistflatsecond}
\end{equation}
see  eqn~(2) in \cite{Meszaros2013}.
The change of variable  
$z=-1+1/a$ in the second definition allows finding the first 
definition.
An analytical expression for the integral
(\ref{lumdistflat}) is here reported 
as a Taylor series of order 8
when  $\om=0.3$ and 
$H_0=70 \h0units$
\begin{eqnarray} 
\dl(z) =
4282.74\,   ( 1+z   )    (  2.45214+ 0.01506\,
   ( 1+z   ) ^{-8}- 0.06633\,   ( 1+z   ) ^{-7}
\nonumber \\
-
 0.01619\,   ( 1+z   ) ^{-6}+ 0.60913\,   ( 1+z
   ) ^{-5}- 1.29912\,   ( 1+z   ) ^{-4}
+ 0.406124\,
   ( 1+z   ) ^{-3}
\nonumber  \\
+ 2.47428\,   ( 1+z   ) ^{-2}-
 4.57509\,   ( 1+z   ) ^{-1}   ) \, Mpc 
\label{dlflat8}
\end{eqnarray}
and the  distance modulus as a function of $z$, $F(z)$, 
\begin{eqnarray} 
(m-M)= F(z)=
43.15861+ 2.17147\,\ln    (  7.77498\,{z}^{2}+ 2.45214
\,{z}^{8}+ 15.0420\,{z}^{7}+ 39.1085\,{z}^{6}
\nonumber \\
+ 56.4947\,{z}^{
5}
+ 49.3673\,{z}^{4}+ 26.1512\,{z}^{3}+ 0.99999\,z+
 1.73 \,10^{-7}  ) - 15.2003\,\ln    ( 1+z   )
\quad .
\label{modulusflat}
\end{eqnarray}
As a consequence, the absolute magnitude, $M$,  is
\begin{equation}
M = m - F(z)
\quad .
\label{absolutemagfz_nonk}
\end{equation} 
The angular 
diameter distance, $D_A$,
after \cite{Etherington1933},
is 
\begin{equation}
D_A = \frac{D_L}{(1+z)^2}
\quad .
\end{equation}
We may  approximate the luminosity distance as given by
eqn~(\ref{dlflat8}) 
by the minimax rational approximation, $d_{L,2,1}$,
with the degree of the numerator $p=2$ and the degree of the denominator $q=1$: 
\begin{equation}
d_{L,2,1} (z) = 
\frac
{
4.10871+ 1813.96\,z+ 2957.04\,{z}^{2}
}
{
0.44404+ 0.27797\,z
}
\label{dlflat21}
\end{equation}
which allows deriving the inverse formula, 
the redshift as a function of the luminosity distance:
\begin{eqnarray}
z_{2,1}(\dl)
=
0.000047\,{\it \dl}
- 0.306718
\nonumber  \\
+
{ 3.38175\times 10^{-14}
}\,\sqrt { 1.9318\,10^{18}\,{{\it \dl}}^{2}+{ 1.06093\times 
10^{23}}\,{\it \dl}+{ 8.10464 \times 10^{25}}}
\quad .
\label{z21}
\end{eqnarray}
Another useful distance is 
the  transverse comoving distance, $D_M$,
\begin{equation} 
D_M = \frac{D_L} {1+z}
\quad ,
\label{comovingdistance}
\end{equation}
with the connected  total comoving volume
$V_c$ 
\begin{equation} 
V_c= \frac{4}{3}\pi D_M^3 
\quad ,
\label{comovingvolume}
\end{equation}
which can be minimax-approximated
as
\begin{equation}
V_{c,3,2} =
\frac 
{
3.01484\,10^{10}{z}^{3}+ 6.39699\,10^{10}\,{z}^{2}- 1.26793\,10^{10}\,z+
 4.10104\,10^8 
}
{
0.45999- 0.01011\,z+ 0.093371\,{z}^{2}
}
\, Mpc^3 \quad .
\end{equation} 

\section{The adopted LFs}
\label{sectruncated}

This section reviews the
Schechter LF, 
the double power law LF, 
and the Pei LF for QSOs.
The truncated version
of the Schechter LF is derived. 
The merit function $\chi^2$
is  computed as
\begin{equation}
\chi^2 =
\sum_{j=1}^n ( \frac {LF_{theo} - LF_{astr} } {\sigma_{LF_{astr}}})^2
\quad ,
\label{chisquare}
\end{equation}
where $n$ is the number of bins for LF of QSOs  and the two 
 indices $theo$ and $astr$ stand for `theoretical' 
and `astronomical', respectively.
The residual sum of squares (RSS) is
\begin{equation}
RSS =
\sum_{j=1}^n ( y(i)_{theo} -y(i)_{astr})^2
\quad ,
\label{rss}
\end{equation}
where  
$y(i)_{theo}$ is the theoretical value
and
$y(i)_{astr}$ is the astronomical value.

A reduced  merit function $\chi_{red}^2$
is  evaluated  by
\begin{equation}
\chi_{red}^2 = \chi^2/NF
\quad,
\label{chisquarereduced}
\end{equation}
where $NF=n-k$ is the number of degrees  of freedom 
and $k$ is the number of parameters.
The goodness  of the fit can be expressed by
the probability $Q$, see  equation 15.2.12  in \cite{press},
which involves the degrees of freedom
and the $\chi^2$.
According to  \cite{press}, the
fit ``may be acceptable'' if  $Q>0.001$.
The Akaike information criterion
(AIC), see \cite{Akaike1974},
is defined by
\begin{equation}
AIC  = 2k - 2  ln(L)
\quad,
\end {equation}
where $L$ is
the likelihood  function  and $k$ is  the number of  free parameters
in the model.
We assume  a Gaussian distribution for  the errors
and  the likelihood  function
can be derived  from the $\chi^2$ statistic
$L \propto \exp (- \frac{\chi^2}{2} ) $
where  $\chi^2$ has been computed by
Equation~(\ref{chisquare}),
see~\cite{Liddle2004}, \cite{Godlowski2005}.
Now the AIC becomes
\begin{equation}
AIC  = 2k + \chi^2
\quad.
\label{AIC}
\end {equation}

\subsection{The Schechter LF}

Let $L$ be a random variable taking
values in the closed interval
$[0, \infty]$.
The Schechter LF of galaxies, 
after  \cite{schechter},
is 
\begin{equation}
\Phi (L;\Phi^*,\alpha,L^*) dL = (\frac {\Phi^*}{L^*}) (\frac {L}{L^*})^{\alpha}
\exp \bigl ( {- \frac {L}{L^*}} \bigr ) dL \quad,
\label{lf_schechter}
\end {equation}
where $\alpha$ sets the slope for low values 
of $L$, 
$L^*$ is the
characteristic luminosity, and $\Phi^*$ represents 
the number of galaxies per $Mpc^3$.
The  normalization is 
\begin{equation}
\int_0^{\infty} \Phi (L;\Phi^*,\alpha,L^*) dL  =
\rm \Phi^*\, \Gamma \left( \alpha+1 \right) 
\quad  , 
\label{norma_schechter}
\end{equation}
where
\begin{equation}
\rm \Gamma \, (z )
=\int_{0}^{\infty}e^{{-t}}t^{{z-1}}dt
\quad ,
\end{equation}
is the gamma function.
The average luminosity,
$ { \langle L \rangle } $, is
\begin{equation}
{ \langle L(\Phi^*,\alpha,L^*) \rangle }
=
\rm L^* \,{\rm \Phi^*  }\,\Gamma \left( \alpha+2 \right) 
\quad  .
\label{ave_schechter}
\end{equation}
An equivalent form  in absolute magnitude 
of the Schechter LF
is  
\begin{equation}
\rm
\Phi (M;\Phi^*,\alpha,M^*)dM=0.921 \Phi^* 10^{0.4(\alpha +1 ) (M^*-M)}
\exp \bigl ({- 10^{0.4(M^*-M)}} \bigr)  dM \, ,
\label{lfstandard}
\end {equation}
where $M^*$ is the characteristic magnitude.
The scaling with  $h$ is  $M^* - 5\log_{10}h$ and 
\\
$\Phi^*h^3~[Mpc^{-3}]$.

\subsection{The truncated Schechter LF}

\label{sectiontruncated}
We assume that the luminosity $L$ takes
values  in the interval
$[L_l, L_u ]$, where the indices $l$ and $u$ mean
`lower' and `upper';
the truncated Schechter   LF, $S_T$, is
\begin {equation}
S_T(L;\Psi^*,\alpha,L^*,L_l,L_u)=
\frac
{
- \left( {\frac {L}{{\it L^*}}} \right) ^{\alpha}{{\rm e}^{-{\frac {
L}{{\it L^*}}}}}{\it \Psi^*}\,\Gamma \left( \alpha+1 \right) 
}
{
{\it L^*}\, \left( \Gamma \left( \alpha+1,{\frac {L_{{u}}}{{\it 
L^*}}} \right) -\Gamma \left( \alpha+1,{\frac {L_{{l}}}{{\it L^*}}
} \right)  \right) 
}
\label{lf_trunc_schechter}
\quad ,
\end{equation}
where $\Gamma(a, z)$ is the incomplete Gamma function
defined as 
\begin{equation}
\mathop{\Gamma\/}\nolimits\!\left(a,z\right)
=\int_{z}^{\infty}t^{a-1}e^{-t}dt 
\quad ,
\end{equation} 
see  \cite{NIST2010}.
The normalization is the same as for the 
Schechter LF,  see eqn~(\ref{norma_schechter}),
\begin{equation}
\int_0^{\infty} S_T(L;\Psi^*,\alpha,L^*,L_l,L_u) dL  =
\rm \Psi^*\, \Gamma \left( \alpha+1 \right) 
\quad . 
\label{norma_schechter_trunc}
\end{equation}
The average value  is  
\begin{equation}
{ \langle L(\Psi^*,\alpha,L^*,L_l,L_u) \rangle }
=
\frac{
N
}
{
{\it L^*}\, \left( \Gamma \left( \alpha+1,{\frac {L_{{u}}}{{\it 
L^*}}} \right) -\Gamma \left( \alpha+1,{\frac {L_{{l}}}{{\it L^*}}
} \right)  \right) 
}
\end{equation}
with 
\begin{eqnarray}
N=
{\it \Psi^*}\, \Bigg  ( {{\it L^*}}^{2}\Gamma \big  ( \alpha+1,{\frac 
{L_{{u}}}{{\it L^*}}} \big  ) \alpha-{{\it L^*}}^{2}\Gamma \big  ( 
\alpha+1,{\frac {L_{{l}}}{{\it L^*}}} \big  ) \alpha+{{\it L^*}}^{
2}\Gamma \big  ( \alpha+1,{\frac {L_{{u}}}{{\it L^*}}} \big  ) 
\nonumber  \\
-{{
\it L^*}}^{2}\Gamma \big  ( \alpha+1,{\frac {L_{{l}}}{{\it L^*}}}
 \big  ) -{{\it L^*}}^{-\alpha+1}{{\rm e}^{-{\frac {L_{{l}}}{{\it 
L^*}}}}}{L_{{l}}}^{\alpha+1}+{{\it L^*}}^{-\alpha+1}{{\rm e}^{-{
\frac {L_{{u}}}{{\it L^*}}}}}{L_{{u}}}^{\alpha+1} \Bigg  ) \Gamma
\big  ( \alpha+1 \big  ) 
\quad  .
\end{eqnarray}
The four luminosities
$L,L_l,L^*$ and $L_u$
are  connected with  the
absolute magnitudes $M$,
$M_l$, $M_u$ and $M^*$
through the following relationship
\begin{equation}
\frac {L}{L_{\sun}} =
10^{0.4(M_{\sun} - M)}
\, ,
\frac {L_l}{L_{\sun}} =
10^{0.4(M_{\sun} - M_u)}
\,
, \frac {L^*}{L_{\sun}} =
10^{0.4(M_{\sun} - M^*)}
\,
, \frac {L_u}{L_{\sun}} =
10^{0.4(M_{\sun} - M_l)}
\label{magnitudes}
\end{equation}
where the indices $u$ and $l$ are inverted in
the transformation
from luminosity to absolute magnitude
and $L_{\sun}$ and  $M_{\sun}$ are  the luminosity and absolute magnitude
of the sun in the considered band.
The equivalent form  in absolute magnitude 
of the truncated Schechter LF is therefore 
\begin{eqnarray}
\rm
\Psi (M;\Psi^*,\alpha,M^*,M_l,M_u)dM
=   \nonumber \\
\frac
{
- 0.4   \left( {10}^{ 0.4  {\it M^*}- 0.4  M} \right) ^{\alpha}{
{\rm e}^{-{10}^{ 0.4  {\it M^*}- 0.4  M}}}{\it \Psi^*}  \Gamma
 \left( \alpha+1 \right) {10}^{ 0.4  {\it M^*}- 0.4  M} \left( \ln 
 \left( 2 \right) +\ln  \left( 5 \right)  \right) 
}
{
\Gamma \left( \alpha+1,{10}^{- 0.4  M_{{l}}+ 0.4  {\it M^*}}
 \right) -\Gamma \left( \alpha+1,{10}^{ 0.4  {\it M^*}- 0.4  M_{{u}
}} \right) 
}
\end{eqnarray}
The averaged absolute magnitude   is  
\begin{equation}
{ \langle M(\Psi^*,\alpha,L^*,L_l,L_u) \rangle }
=
\frac{
\int_{M_l}^{M_u} M(M;\Psi^*,\alpha,L^*,L_l,L_u) M 
dM
}
{
\int_{M_l}^{M_u} M(M;\Psi^*,\alpha,L^*,L_l,L_u) 
dM 
}
\quad  .
\label{xmtruncated}
\end{equation}

\subsection{The double power law}

The double power law LF for QSOs 
is 
\begin{equation}
  \Phi(L;\phi^*,\alpha,\beta,L^*) = \frac{ \phi^*}
                            {  (L/L^{*})^{\alpha}    +  (L/L^{*})^{\beta}  }
\quad ,
 \label{doublepowerlaw}
\end{equation}
where
$L^{*}$   is the characteristic   luminosity,
$\alpha$  models  the low boundary, 
and
$\beta$  models  the high boundary,
 see 
\cite{Boyle1988,Boyle2000,Croom2004,Richards2006,Ross2013,Singh2016c}.
The magnitude version  is 
\begin{equation}
  \Phi(M;\phi^*,\alpha,\beta,M^*) = \frac{ \phi^*}
       { 10^{0.4{(\alpha +1)[M-M^{*}]}}+10^{0.4{(\beta +1)[M-M^{*} ]}} }
 \label{doublepowerlawmag}
\quad ,
\end{equation}
where the characteristic absolute magnitude, $M^{*}$,
and $\phi^*$ are  
functions  of the redshift.

\subsection{The Pei function}

The  exponential  $L^{1/4}$ LF, or Pei LF, after \cite{Pei1995}, 
is
\begin{equation}
 \Phi(L;\phi^*,\beta,L^*) =
\frac
{
{\it \phi^*}\, \left( {\frac {L}{{\it L^*}}} \right) ^{-\beta}{
{\rm e}^{-\sqrt [4]{{\frac {L}{{\it L^*}}}}}}
}
{
{\it L^*}
}
\quad  ,
\end{equation} 
and the magnitude version is 
\begin{equation}
  \Phi(M;\phi^*,\beta,M^*) = 
\frac
{
0.4\,{\it \phi^*}\, \left( {\frac {{10}^{- 0.4\,M}}{{10}^{- 0.4\,{M}^
{{\it *}}}}} \right) ^{-\beta}{{\rm e}^{-\sqrt [4]{{\frac {{10}^{-
 0.4\,M}}{{10}^{- 0.4\,{M}^{{\it *}}}}}}}}{10}^{- 0.4\,M}\ln 
 \left( 10 \right) 
}
{
{10}^{- 0.4\,{M}^{{\it *}}}
}
 \label{pei14mag}
\quad .
\end{equation}

\section{The astrophysical applications}
\label{secqso}

This section 
explains the K-correction for QSOs,
introduces 
the sample of QSOs on which 
the various tests are performed,
finds the parameters of the new LF 
in the range of redshift $[0.3 , 0.5]$,
and finds the number of QSOs  as a function of
the redshift.

\subsection{K-correction}

The K-correction for QSOs  as 
f unction of the redshift  can be  parametrized 
as  
\begin{equation}
K(z) = -2.5 \,  (1+\alpha_{\nu}) \log (1+z)
\quad ,
\end{equation}
with  $-0.7 < \alpha_{\nu} < -0.3$, see \cite{Wisotzki2000}. 
Following \cite{Croom2009}, we have adopted $\alpha_{\nu}=-0.3$.
The corrected absolute magnitude, $M_K$, is 
\begin{equation}
M_K= M + K(z)
\quad .
\label{absolutemagfz}
\end{equation}
In the following, both the observed and the theoretical 
absolute magnitude will always be K-corrected.

\subsection{The sample of QSO}

\begin{figure}
\begin{center}
\includegraphics[width=7cm]{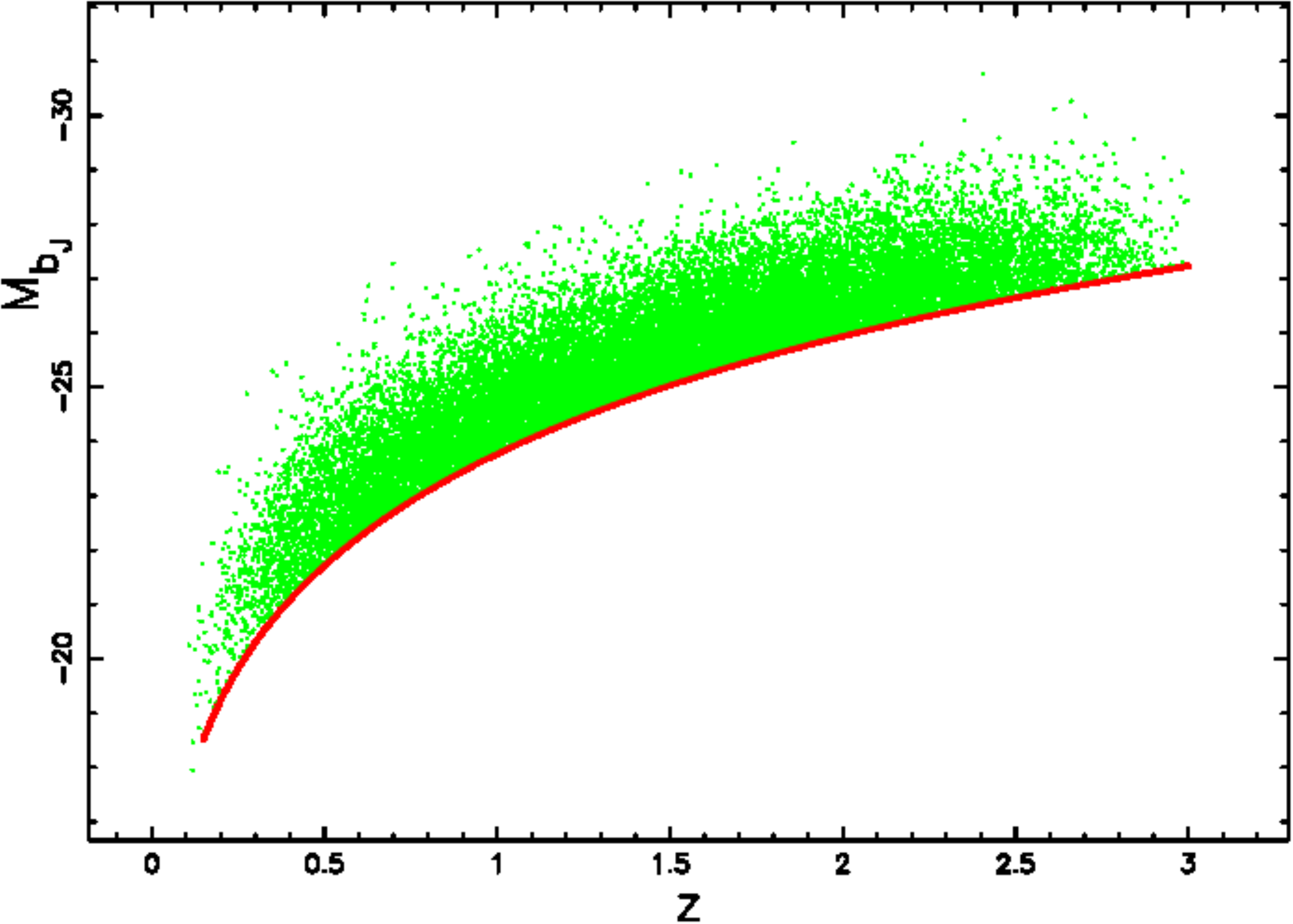}
\end{center}
\caption{
The absolute magnitude $M_{B_{bj}}$
computed with the nonlinear Eq.~(\ref{absolutemagfz})
for
22413 QSOs versus the  redshift, (green points).
The lower theoretical curve (upper absolute magnitude)
as represented by
the nonlinear Eq.~(\ref{absolutemagfzupper}) is  the
red thick line.
The redshifts cover the range $[0,3]$
}
 \label{qsolowermag}%
\end{figure}

We selected the  catalog of the 2dF QSO Redshift Survey (2QZ),
which contains  22431 redshifts of QSOs  
with $18.25 <b_J< 20.85$,
a total survey area of    721.6 $\deg^2$,
and an effective area of  673.4 $\deg^2$, 
see \cite{Croom2004} 
\footnote{
Data at 
\url{http://vizier.u-strasbg.fr/viz-bin/VizieR?-source=VII/241}.
}
.
Section 3 in \cite{Croom2004}  
discusses  four separate types of completeness
which characterize 
the 2QZ and 6QZ surveys:          
 (i)  morphological completeness, $f_m(bJ,z)$,
 (ii) photometric completeness,   $f_p(bJ, z)$,
 (iii)coverage completeness       $f_c(\theta)$ and
 (iv) spectroscopic completeness, $f_s(bJ, z, \thetaθ)$.
The first test can be done  
on the  upper limit of  the maximum
absolute  magnitude,
$M_u(z)$,
 which  can be observed in a
catalog of QSOs  characterized by a given limiting
magnitude, in our case $b_j=20.85$,
where $F(z)$ has  been defined
by eqn~(\ref{modulusflat}):
\begin{equation}
M_u(z) = 20.85 - F(z)
\quad ,
\label{absolutemagfzupper}
\end{equation} 
see Figure \ref{qsolowermag}. 

%begin figure qsolowermag
% end figure qsolowermag
A careful examination of Figure  \ref{qsolowermag} 
allows concluding that all
the QSOs are in the region over the border line,
the number of observed QSOs
decreases with increasing $z$, and 
the average absolute magnitude decreases with increasing $z$.
The previous comments  can be connected with the
 Malmquist bias,
see \cite{Malmquist_1920,Malmquist_1922},
which was originally applied
to the stars and later on
to the galaxies by \cite{Behr1951}.

\subsection{The luminosity function for QSOs}

A binned luminosity function for quasars can be built 
in one of the two methods suggested by \cite{Page2000}: 
the $\frac{1}{V_a}$ method , see \cite{Avni1980,Eales1993,Ellis1996}, 
and a binned approximation.
Notably, \cite{Yuan2013},  argued that
both the $\frac{1}{V_a}$ and the binned approximation 
can produce bias at the faint
end of the LF due to the
arbitrary choosing of redshift and luminosity intervals.

We implemented the binned approximation
of  \cite{Page2000}, $\phi_{est}$,
as 
\begin{equation}
\label{lfapprox}
\phi \approx \phi_{est} = \frac{N_q}{\int_{M_{min}}^{M_{max}}
\int^{z_{max}(M)}_{z_{min}}
\frac{dV}{dz} dz dM}
\quad ,
\end{equation}
where $N_q$ is the number of quasars observed in the $M_i-z$ bin.
The error is evaluated as 
\begin{equation}
\label{lfapprox_error}
\delta \phi_{est} = \frac{\sqrt{N_q}}{\int_{M_{min}}^{M_{max}}
\int^{z_{max}(M)}_{z_{min}}
\frac{dV}{dz} dz dM}
\quad .
\end{equation}
The comoving volume  in the flat cosmology is evaluated 
according to equation (\ref{comovingvolume}),
\begin{equation}
V =\frac{4}{3} \pi (D_{M,upp}^3 - D_{M,low}^3) 
\quad ,
\end{equation}
where  $D_{M,upp}$ and  $D_{M,low}$ are, 
respectively,
the upper and lower comoving  distance.
A correction for  the effective 
volume of the catalog, $V_q$,
gives 
\begin{equation}
V_q = V \frac{A_e \, deg^2}{41252.9\, deg^2} \quad ,
\end{equation}
where $A_e$ is the effective area of the catalog in $deg^2$.

\begin{figure}
\begin{center}
\includegraphics[width=7cm]{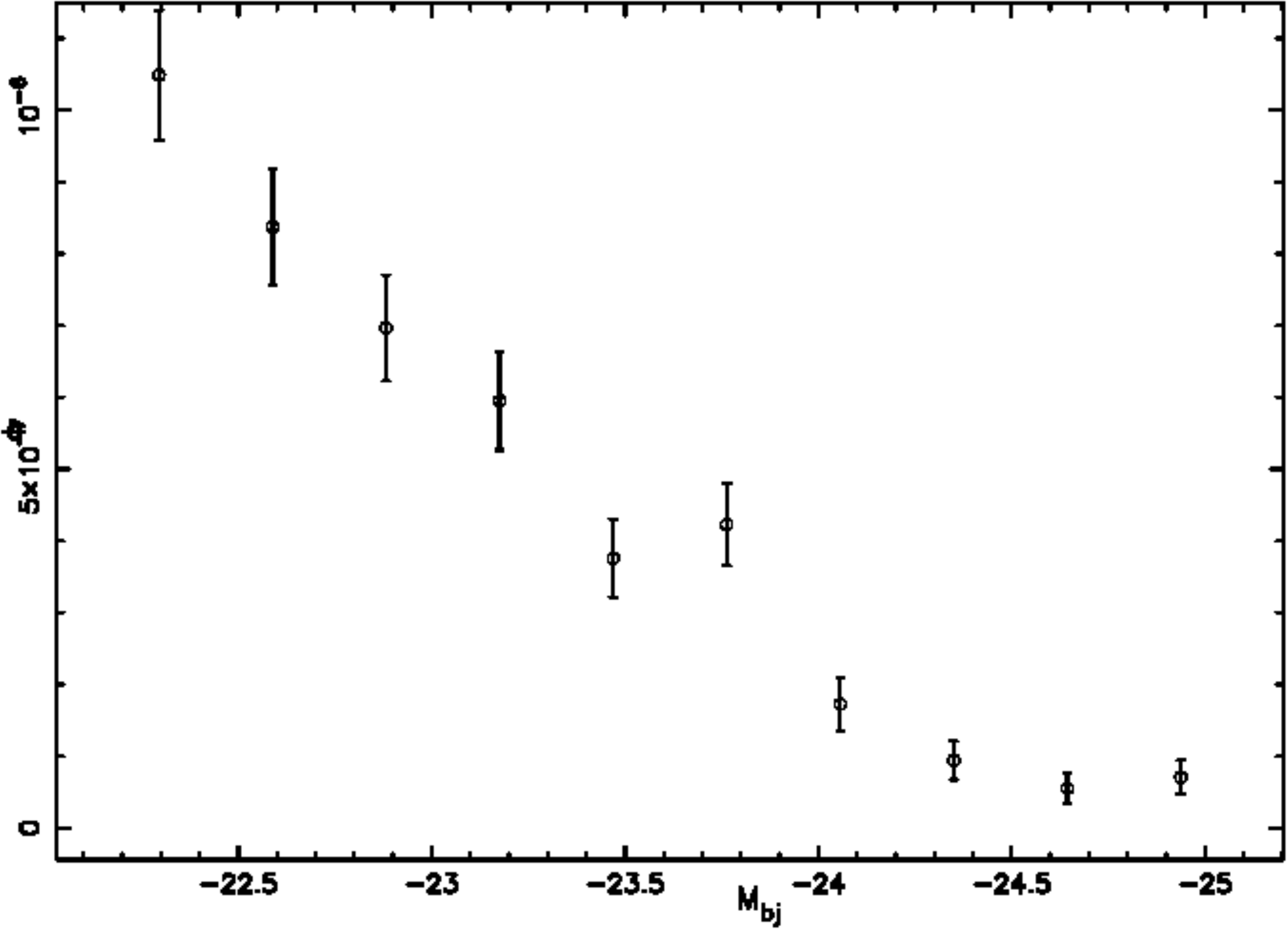}
\end{center}
\caption{
The  observed LF for QSOs
is  reported with the error bar evaluated 
as the square root of  the LF (Poissonian distribution)
when $z$  $[0.3,0.5]$.  
}
 \label{lfqso03lf}%
\end{figure}

\begin{figure}
\begin{center}
\includegraphics[width=7cm]{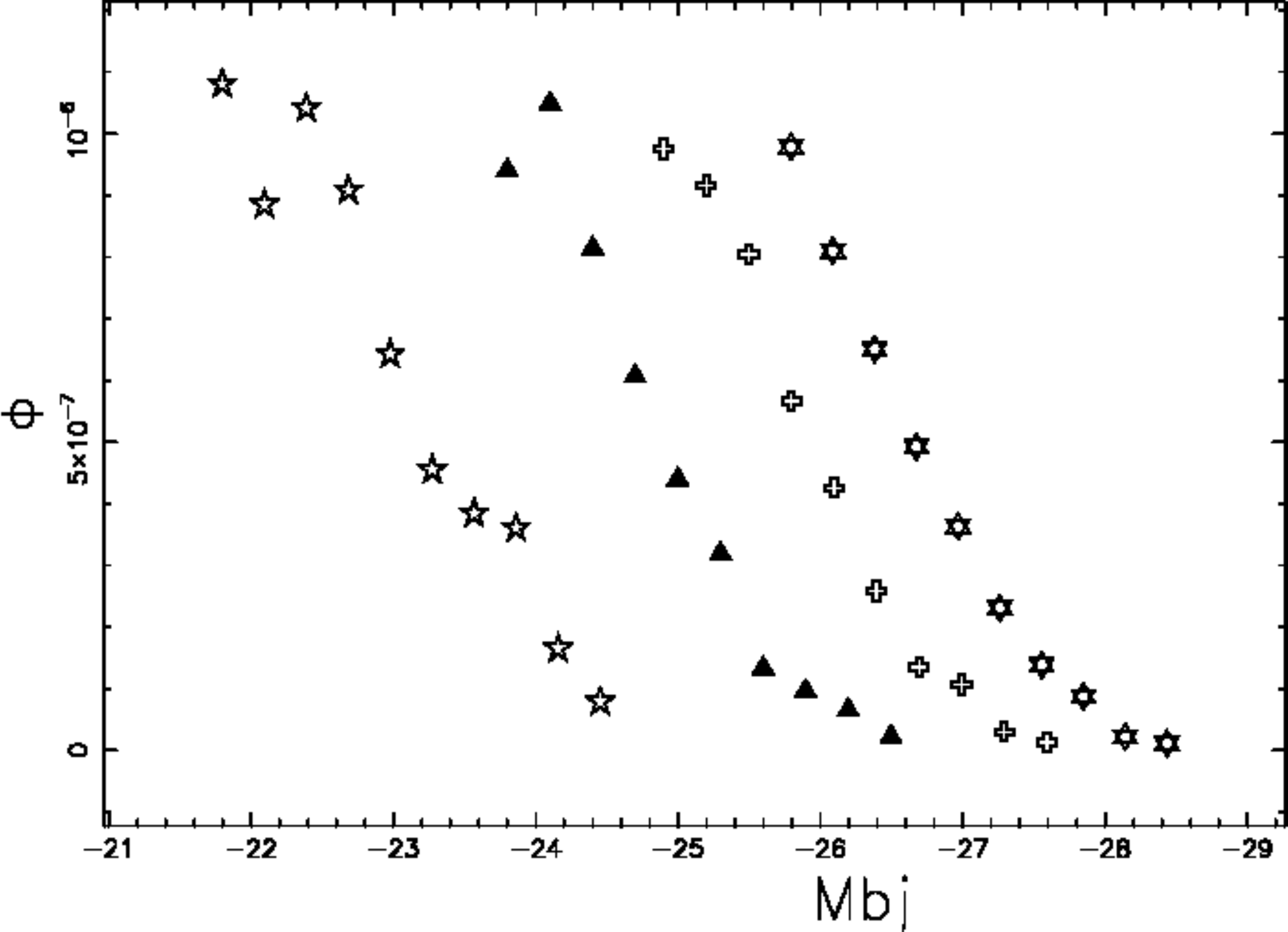}
\end{center}
\caption{
The  observed LF for QSOs
when $z$   
$[0.3,0.5]$  and M $[-24.45,-21.50]$ (empty stars),
$[0.7,0.9]$  and M $[-26.49,-23.50]$ (full triangles),
$[1.1,1.3]$  and M $[-27.59,-24.60]$ (empty crosses ) and
$[1.5,1.7]$  and M $[-28.43,-25.50]$ (stars of David).
}
 \label{lfqso_piu}%
\end{figure}

A typical  example of the observed LF for QSOs 
when  $ 0.3 < z< 0.5$ is reported
in Figure \ref{lfqso03lf}    
%begin figure lfqso03lf
% end figure lfqso03lf
and Figure \ref{lfqso_piu} reports 
the LF for QSOs in four ranges of redshift.

%begin figure lfqso_piu
% end figure lfqso_piu
%siamoqui  
The   variable  lower  bound in absolute magnitude, 
$M_l$  can  be  connected  with  evolutionary effects,
and 
the upper bound, $M_u$, is fixed by the physics, see  
the nonlinear Eq.~(\ref{absolutemagfzupper}),
see Section \ref{secevolutionary}. 

\begin{figure}
\begin{center}
\includegraphics[width=7cm]{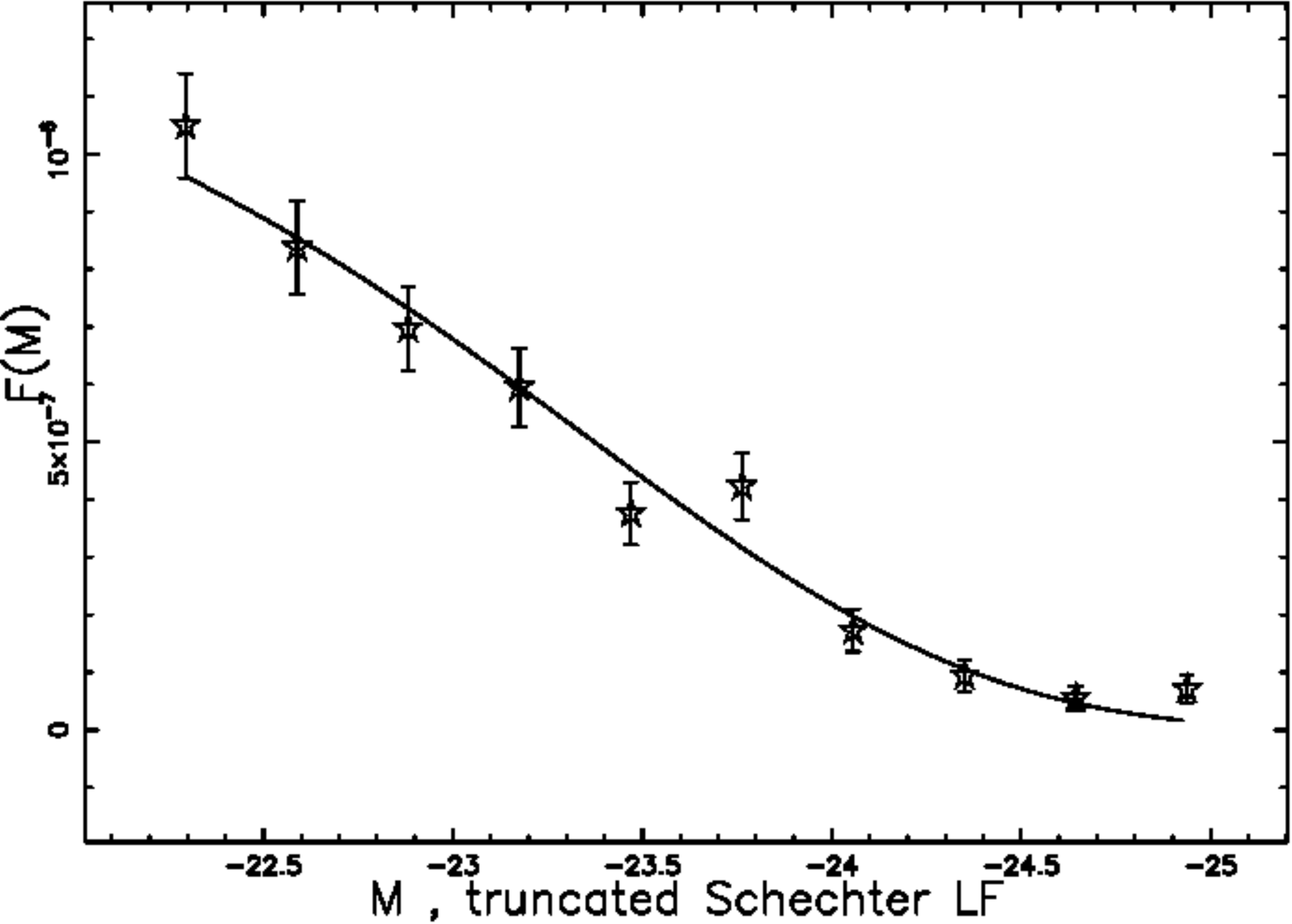}
\end{center}
\caption{
The  observed LF for QSOs, empty stars with error bar, 
and the fit  by  the truncated Schechter LF
when $z$ $[0.3,0.5]$
and  $M$ $[-24.93,-22]$.
}
 \label{trschechterlfqso03}%
\end{figure}

The five parameters  of the 
the best fit to the observed LF 
by  the truncated Schechter LF
can be found with the Levenberg--Marquardt method and 
are    reported in 
Table \ref{trschechterfit}.
The resulting fitted curve
is displayed in Figure  \ref{trschechterlfqso03}.
%begin figure trschechterlfqso03
% end figure trschechterlfqso03

\begin{table}
\caption
{
Parameters  of the  truncated Schechter LF 
in the range of redshifts  $[0.3,0.5]$
when n=10 and k=5.
}
 \label{trschechterfit}
 \[
 \begin{array}{ccccccccc}
 \hline
 \hline
 \noalign{\smallskip}
M_l   
&M^* 
&  M_u~ ,
& \Psi^* 
&  \alpha 
& \chi^2
& \chi_{red}^2
& Q  
& AIC  
\\
 \noalign{\smallskip}
 \hline
 -24.93 & -23.28   & -22.29 & 3.38\,10^{-8} &  -0.97 & 12.89 
& 2.57 & 0.024 & 22.89 \\
 \hline
 \hline
 \end{array}
 \]
\end{table}

For  the sake of comparison, Table \ref{schechterfit}
reports the three parameters of the  Schechter LF.

\begin{table}
\caption
{
Parameters  of the  Schechter LF 
in the range $[0.3,0.5]$
when k=3 and n=10.
}
 \label{schechterfit}
 \[
 \begin{array}{ccccccc}
 \hline
 \hline
 \noalign{\smallskip}
 M^*
& \Psi^*
&  \alpha 
&  \chi^2
& \chi_{red}^2
& Q  
& AIC  
\\
 \noalign{\smallskip}
 \hline
 -23.75  & 8.85\,10^{-7}& -1.37 & 10.49 
& 1.49  & 0.162  & 16.49
\\
 \hline
 \hline
 \end{array}
 \]
\end{table}

\begin{figure}
\begin{center}
\includegraphics[width=7cm]{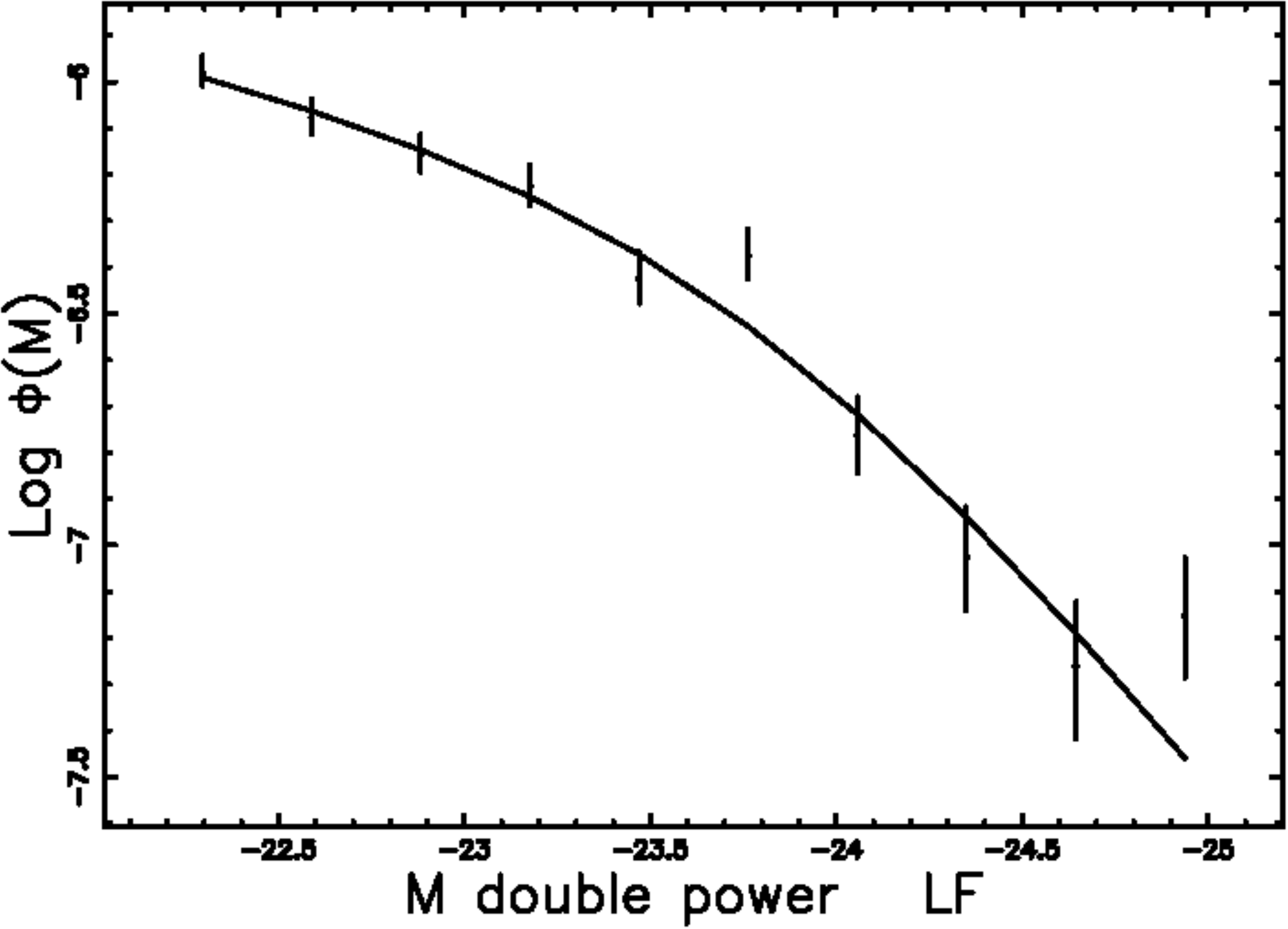}
\end{center}
\caption{
The  observed LF for QSOs, empty stars with error bar, 
and the fit  by  the double power  LF
when the redshifts cover the range $[0.3,0.5]$
}
 \label{double_power03}%
\end{figure}

As a {\it first} reference 
the  fit with the double power LF, see equation (\ref{doublepowerlawmag}), 
is displayed in Figure  \ref{double_power03}
with  parameters as in 
Table  \ref{doublepowerfit}. 

%begin figure double_power03
% end figure double_power03

\begin{table}
\caption
{
Parameters  of the  double power LF 
in the range of redshifts  $[0.3,0.5]$
when n=10 and k=4.
}
 \label{doublepowerfit}
 \[
 \begin{array}{cccccccc}
 \hline
 \hline
 \noalign{\smallskip}
M^*
&  \phi^*
&  \alpha 
&  \beta 
&  \chi^2 
& \chi_{red}^2
& Q  
& AIC  
\\
 \noalign{\smallskip}
 \hline
-23.82 & 5.44\,10^{-7} & -3.57 & -1.48 & 9.44 
& 1.57   & 0.15  & 17.44
\\
 \hline
 \hline  
 \end{array}
 \]
\end{table}

\begin{figure}
\begin{center}
\includegraphics[width=7cm]{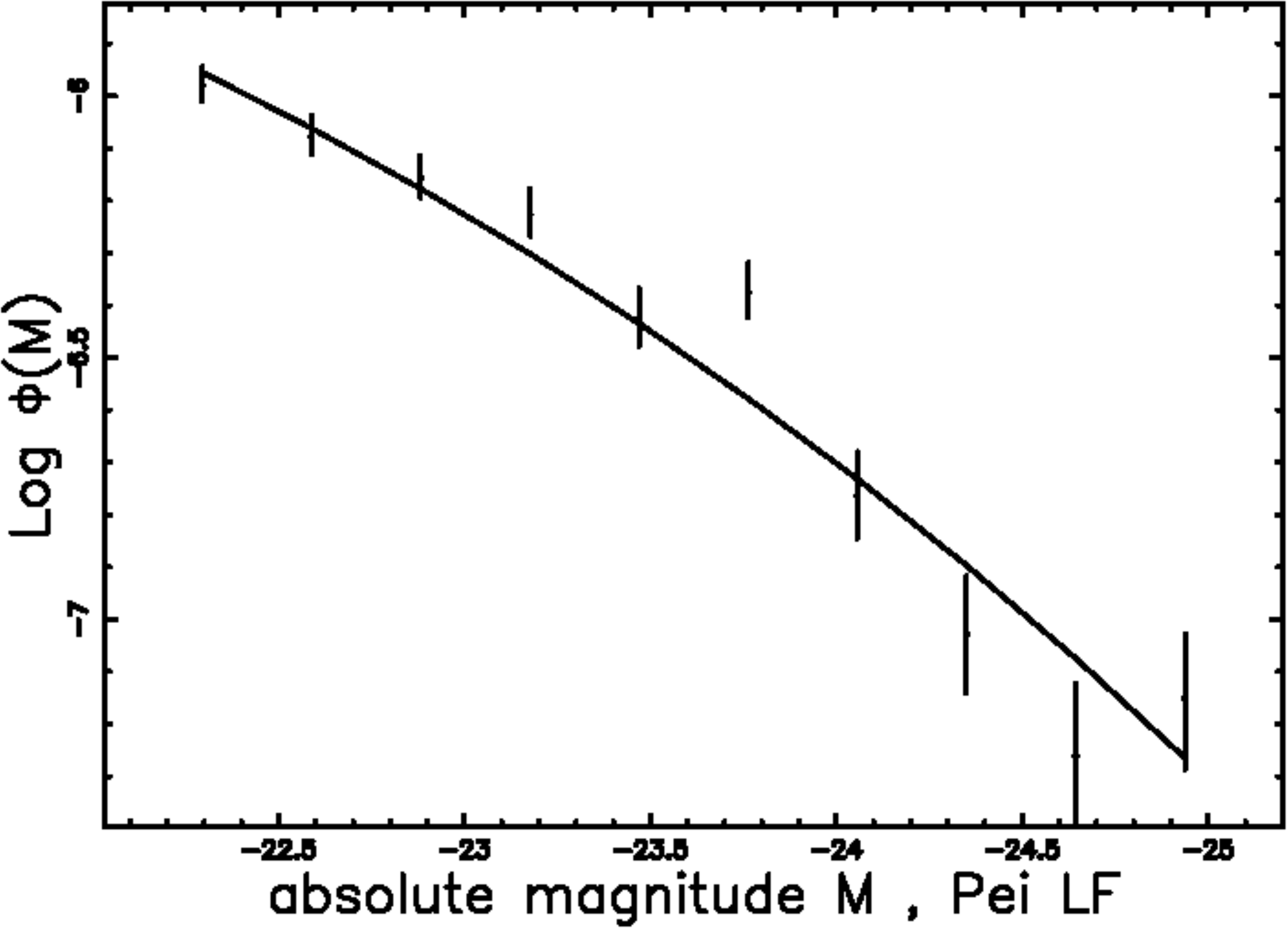}
\end{center}
\caption{
The  observed LF for QSOs, empty stars with error bar, 
and the fit  by  the Pei    LF
when the redshifts cover the range $[0.3,0.5]$
}
 \label{pei14_03}%
\end{figure}

As a {\it second} reference 
the  fit with the Pei  LF, see equation (\ref{pei14mag}), 
is displayed in Figure  \ref{pei14_03}
with  parameters as in 
Table  \ref{doublepowerfit}. 
%begin figure pei14_03
% end figure pei14_03

\begin{table}
\caption
{
Parameters  of the Pei  LF 
in the range of redshifts  $[0.3,0.5]$
with k=3 and n=10.
}
 \label{pei14fit}
 \[
 \begin{array}{ccccccc}
 \hline
 \hline
 \noalign{\smallskip}
M^* 
 &  \phi^*
 &  \beta 
 &  \chi^2
 & \chi_{red}^2
 & Q  
 & AIC  
 \\
 \noalign{\smallskip}
 \hline
-16.47 & 3.68\,10^{-5}  & 0.924 & 14.4 
&2.05   &  0.044   & 20.40
\\
 \hline
 \hline
 \end{array}
 \]
\end{table}

\subsection{Evolutionary effects}

\label{secevolutionary} 
\begin{figure}
\begin{center}
\includegraphics[width=10cm]{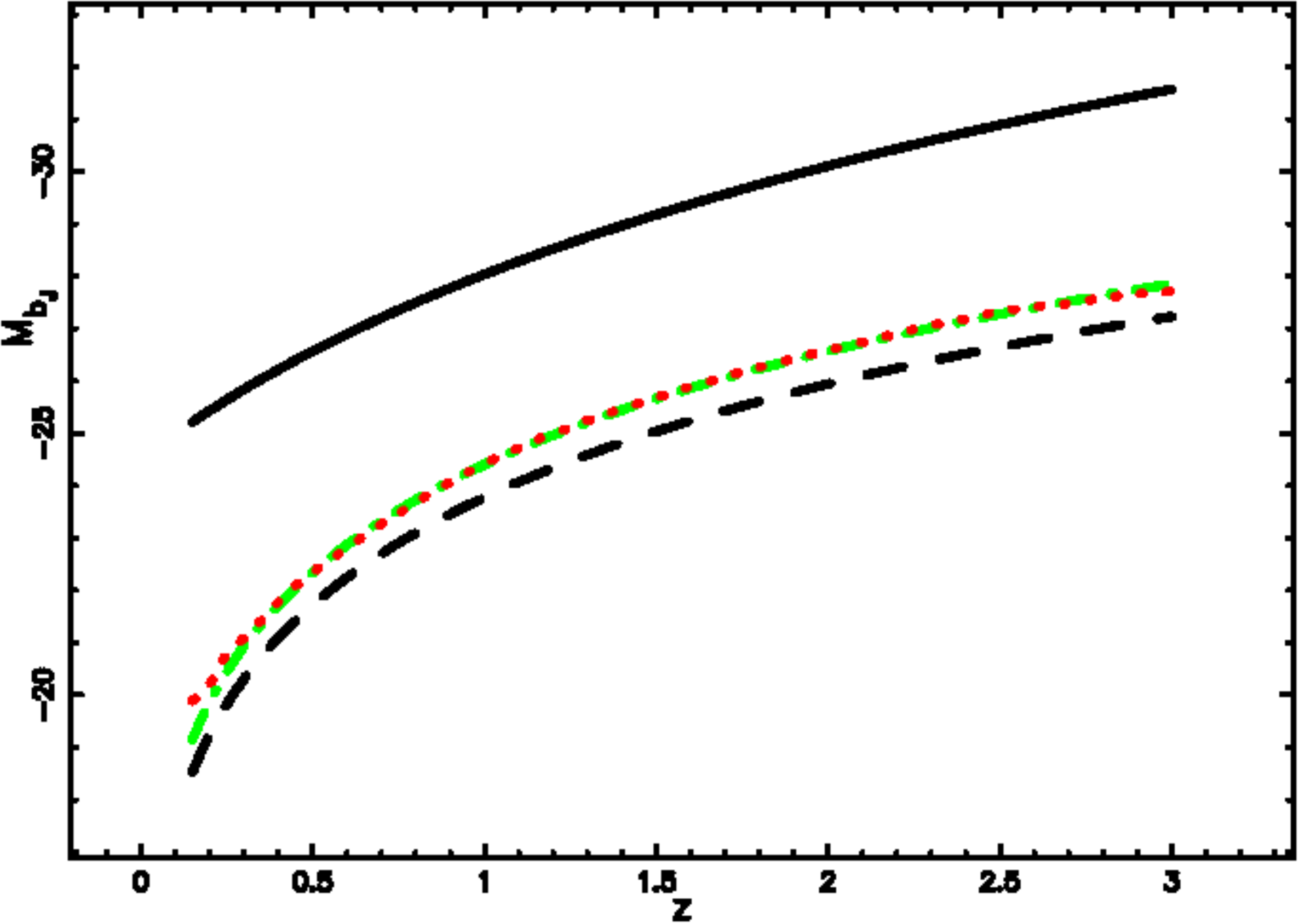}
\end{center}
\caption
{
Average observed  absolute magnitude 
versus redshift for QSOs  (red points),
average theoretical absolute magnitude 
for truncated Schechter   LF
as given by eqn~(\ref{xmtruncated})
(dot-dash-dot green line),
theoretical curve for the empirical 
lowest absolute magnitude  at  a given
redshift,
see eqn~(\ref{mlz}) (full black line) and
the theoretical  curve
for the highest absolute magnitude  at  a given
redshift (dashed black line),
see eqn~(\ref{absolutemagfzupper}),
RSS=1.212.
}
\label{qsoxmz}
\end{figure}

In order to model the evolutionary effects,
an empirical  variable  lower  bound in absolute magnitude, 
$M_l$,
has been introduced,
\begin{equation}
M_l(z)  =-24.5  - 10 \times \log_{10}(1+z) + K(z)  
\label{mlz}
\quad .
\end{equation}
The above empirical formula 
is classified as top line in Figure 5 of \cite{Croom2009} 
and  connected with the limits in magnitude.
Conversely
the upper bound, $M_u$  was already fixed
by the nonlinear Eq.~(\ref{absolutemagfzupper}).
A second evolutionary correction  is
\begin{equation}
M^*= M_u(z)-0.5
\quad ,
\label{mstarcorrection}
\end{equation}
where  
$M_u(z)$  has  been defined in 
eqn~(\ref{absolutemagfzupper}).
Figure \ref{qsoxmz}
reports a comparison between
the theoretical and the observed average absolute magnitudes;
the value of $M^*$ reported in eqn~(\ref{mstarcorrection})
minimizes the difference between the two curves.
% figure   qsoxmz
% end qsoxmz

\begin{figure}
\begin{center}
\includegraphics[width=10cm]{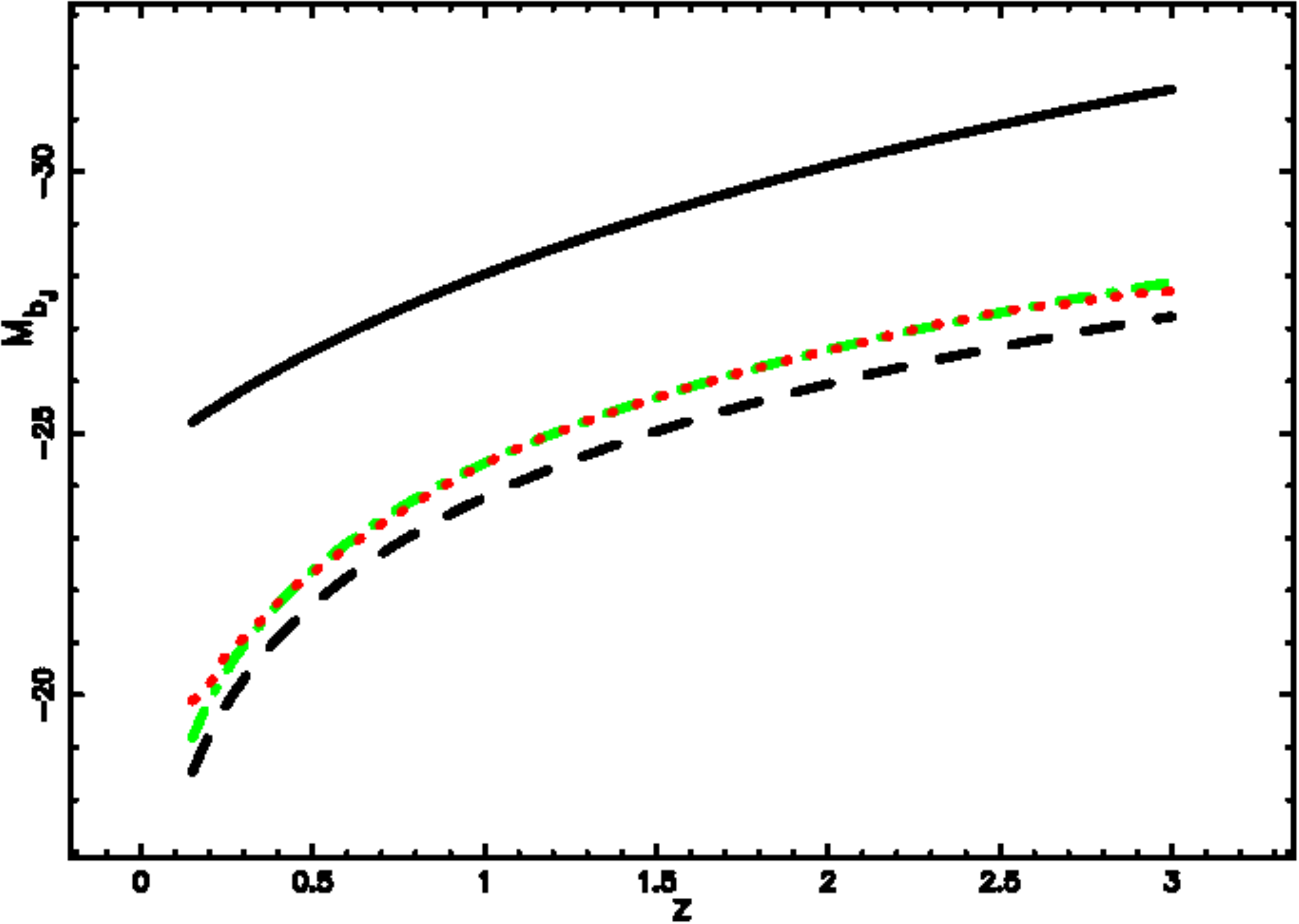}
\end{center}
\caption
{
Average observed  absolute magnitude 
versus redshift for QSOs  (red points),
average theoretical absolute magnitude 
for the double power    LF
as  evaluated numerically 
(dot-dash-dot green line),
theoretical curve for the empirical 
lowest absolute magnitude  at  a given
redshift,
see eqn~(\ref{mlz}) (full black line) and
the theoretical  curve
for the highest absolute magnitude  at  a given
redshift (dashed black line),
see eqn~(\ref{absolutemagfzupper}),
RSS= 1.138.
}
\label{qsoxmz_double}
\end{figure}

As a {\it first} reference Figure \ref{qsoxmz_double}
reports a comparison between
the theoretical and the observed average absolute magnitudes
in the case of the double power LF;
the value of $M^*$ which 
minimizes the difference between the two curves
\begin{equation}
M^*= M_u(z)-0.4
\quad ,
\label{mstarcorrection_double}
\end{equation}
and other parameters as in Table \ref{doublepowerfit}.

% figure   qsoxmz_double
% end qsoxmz_double

\begin{figure}
\begin{center}
\includegraphics[width=10cm]{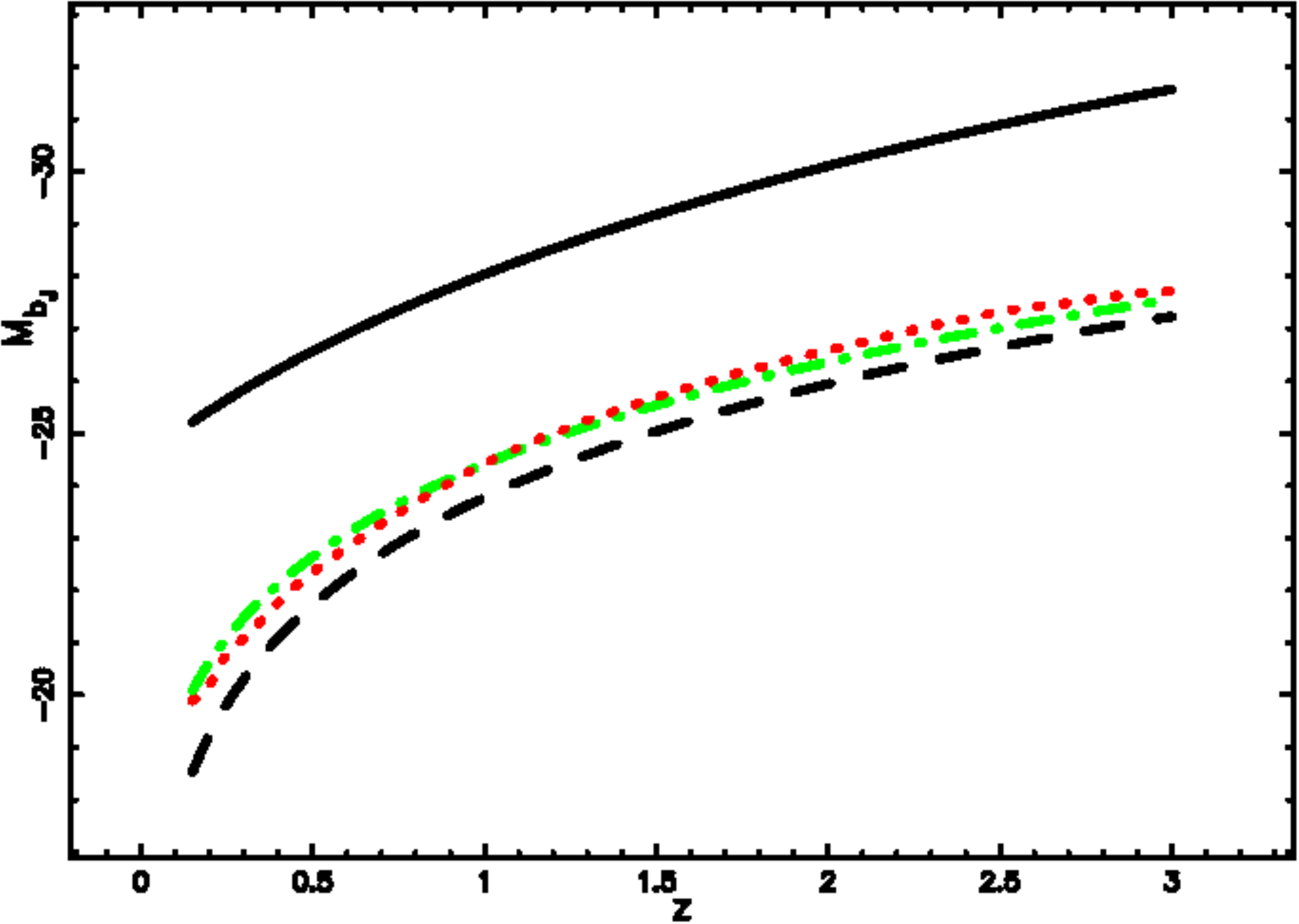}
\end{center}
\caption
{
Average observed  absolute magnitude 
versus redshift for QSOs  (red points),
average theoretical absolute magnitude 
for the Pei   LF
as  evaluated numerically 
(dot-dash-dot green line),
theoretical curve for the empirical 
lowest absolute magnitude  at  a given
redshift,
see eqn~(\ref{mlz}) (full black line) and
the theoretical  curve
for the highest absolute magnitude  at  a given
redshift (dashed black line),
see eqn~(\ref{absolutemagfzupper}),
RSS= 5.41.
}
\label{qsoxmz_pei14}
\end{figure}

As a {\it second } reference Figure \ref{qsoxmz_pei14}
reports a comparison between
the theoretical and the observed average absolute magnitude
in the case of the Pei LF 
with  parameters as in Table \ref{pei14fit}.

% figure   qsoxmz_pei14
% end qsoxmz_pei14
In the above fit, the evolutionary correction for $M^*$ 
is absent.

\subsection{The photometric maximum}

The definition of the flux,$f$,
is 
\begin{equation}
f  = \frac{L}{4 \pi r^2}
\quad ,
\label{flux}
\end{equation} 
where $r$ is the luminosity  distance.
The redshift  is approximated as 
\begin{equation}
 z =z_{2,1}
\quad ,
\end{equation}
where  $z_{2,1}$ has  been introduced
into eqn~(\ref{z21}).
 The relation between $dr$
 and $dz$ is
\begin{equation}
dr =  \frac
{
\left(  2626.1\,z+ 821.99\,{z}^{2}+ 804.33 \right) {
}
}
{
\left(  0.44404+ 0.27797\,z \right) ^{2}
}
dz 
\quad  ,
\end{equation}
where $r$ has been defined as $d_{L,2,1}$
 by the  minimax rational approximation,
see eqn~(\ref{dlflat21}).
The joint distribution in {\it z}
and  {\it f}  for the number of galaxies
 is
\begin{equation}
\frac{dN}{d\Omega dz df} =
\frac{1}{4\pi}\int_0^{\infty} 4 \pi r^2 dr 
S_T(L;\Psi^*,\alpha,L^*,L_l,L_u)
\delta\bigl(z- (z_{2,1})\bigr)
\delta\bigl(f-\frac{L}{4 \pi r^2}    \bigr)
\quad ,
\label{nfunctionzsscztrunc}
\end{equation}
where $\delta$ is the Dirac delta function
and  $S_T(L;\Psi^*,\alpha,L^*,L_l,L_u)$  has  been defined
in eqn~(\ref{lf_trunc_schechter}).  
The above formula has the following explicit version
\begin{equation}
\frac{dN}{d\Omega dz df} = \frac {NL}{DL}
\quad ,
\end{equation}
where 
\begin{eqnarray}
NL = -{ 1.71174\times 10^{21}}  \left( z+ 0.61116 \right) ^{4}
 \left( z+ 0.00227 \right) ^{4}  \times
\nonumber \\
\left(  1.422\,10^9 {\frac {f
 \left( z+ 0.61116 \right) ^{2} \left( z+ 0.00227 \right) 
^{2}}{ \left( z+ 1.59739 \right) ^{2}{\it L^*}}} \right) ^{
\alpha}{{\rm e}^{- 1.422\,10^9 {\frac {f \left( z+ 0.61116
 \right) ^{2} \left( z+ 0.00227 \right) ^{2}}{ \left( z+
 1.59739 \right) ^{2}{\it L^*}}}}} \times 
\nonumber  \\
{\it \Psi^*} \Gamma \left( 
\alpha+ 1  \right)  \left( z+ 2.85165 \right)  \left( z+
 0.343138 \right) 
\end{eqnarray}
where 
\begin{eqnarray}
DL =
\left( z+ 1.59739 \right) ^{6}{\it L^*}\, \left( \Gamma \left( 
\alpha+ 1,{\frac {{\it L_u}}{{\it L^*}}} \right) - \Gamma
 \left( \alpha+ 1,{\frac {{\it L_l}}{{\it L^*}}} \right)  \right) 
\quad .
\end{eqnarray}
The magnitude version is
\begin{eqnarray}
\frac{dN}{d\Omega dz dm}=
\frac{NM}{DM}
\quad ,
\label{nzapparent}
\end{eqnarray}
with 
\begin{eqnarray}
NM  =
-{ 1.25459\times 10^{30}}\, \left( z+ 0.61116 \right) ^{4}
 \left( z+ 0.00227 \right) ^{4} 
\times \nonumber \\
\left(  1.13159\times 10^{18}\,
{\frac {{{\rm e}^{ 0.92103\,{\it M_{\sun}}- 0.92103\,{\it 
m}}} \left( z+ 0.61116 \right) ^{2} \left( z+
 0.00227 \right) ^{2}}{ \left( z+ 1.59739 \right) ^{2}{10}^
{ 0.4\,{\it M_{\sun}}- 0.4\,{\it M^*}}}} \right) ^{\alpha}
\times\nonumber \\
{{\rm e}^{-
 1.13159\times 10^{18}\,{\frac {{{\rm e}^{ 0.92103\,{\it M_{\sun}}-
 0.92103\,{\it m}}} \left( z+ 0.61116 \right) ^{2}
 \left( z+ 0.00227 \right) ^{2}}{ \left( z+ 1.59739
 \right) ^{2}{10}^{ 0.4\,{\it M_{\sun}}- 0.4\,{\it M^*}}}}}}
\times\nonumber \\
{\it 
\Psi^*}\,\Gamma \left( \alpha+ 1.0 \right)  \left( z+ 2.85165
 \right)  \left( z+ 0.34313 \right) {{\rm e}^{ 0.92103\,{
\it M_{\sun}}- 0.92103\,{\it m}}}
\end{eqnarray}
and
\begin{eqnarray}
DM =
\left( z+ 1.59739 \right) ^{6}{10}^{ 0.4\,{\it M_{\sun}}- 0.4\,{
\it M^*}}
\times  
\nonumber \\
\left( \Gamma \left( \alpha+ 1,{\frac {{10}^{ 0.4\,{
\it M_{\sun}}- 0.4\,{\it M_l}}}{{10}^{ 0.4\,{\it M_{\sun}}- 0.4\,{\it 
M^*}}}} \right) - \Gamma \left( \alpha+ 1,{\frac {{10}^{ 0.4
\,{\it M_{\sun}}- 0.4\,{\it M_u}}}{{10}^{ 0.4\,{\it M_{\sun}}- 0.4\,{\it 
M^*}}}} \right)  \right) 
\quad , 
\end{eqnarray}
where $m$ is the  apparent magnitude of the catalog,
the absolute magnitudes 
$M_l$, $M_u$,$M^*$ and $M_{\sun}$ have 
been defined in Section \ref{sectiontruncated}.
The conversion from flux, $f$, to apparent magnitude, $m$,
in the above formula is obtained from  the usual formula
\begin{equation}
f=7.95774\times10^8\,{{\rm e}^{ 0.92103\,{\it M_{\sun}}- 0.92103\,{
\it m}}}
\quad ,
\end{equation}
and
\begin{equation}
df=- 7.32935\times 10^8\,{{\rm e}^{ 0.92103\,{\it M_{\sun}}- 0.92103\,m}
} dm 
\quad .
\end{equation}

\begin{figure}
\begin{center}
\includegraphics[width=6cm]{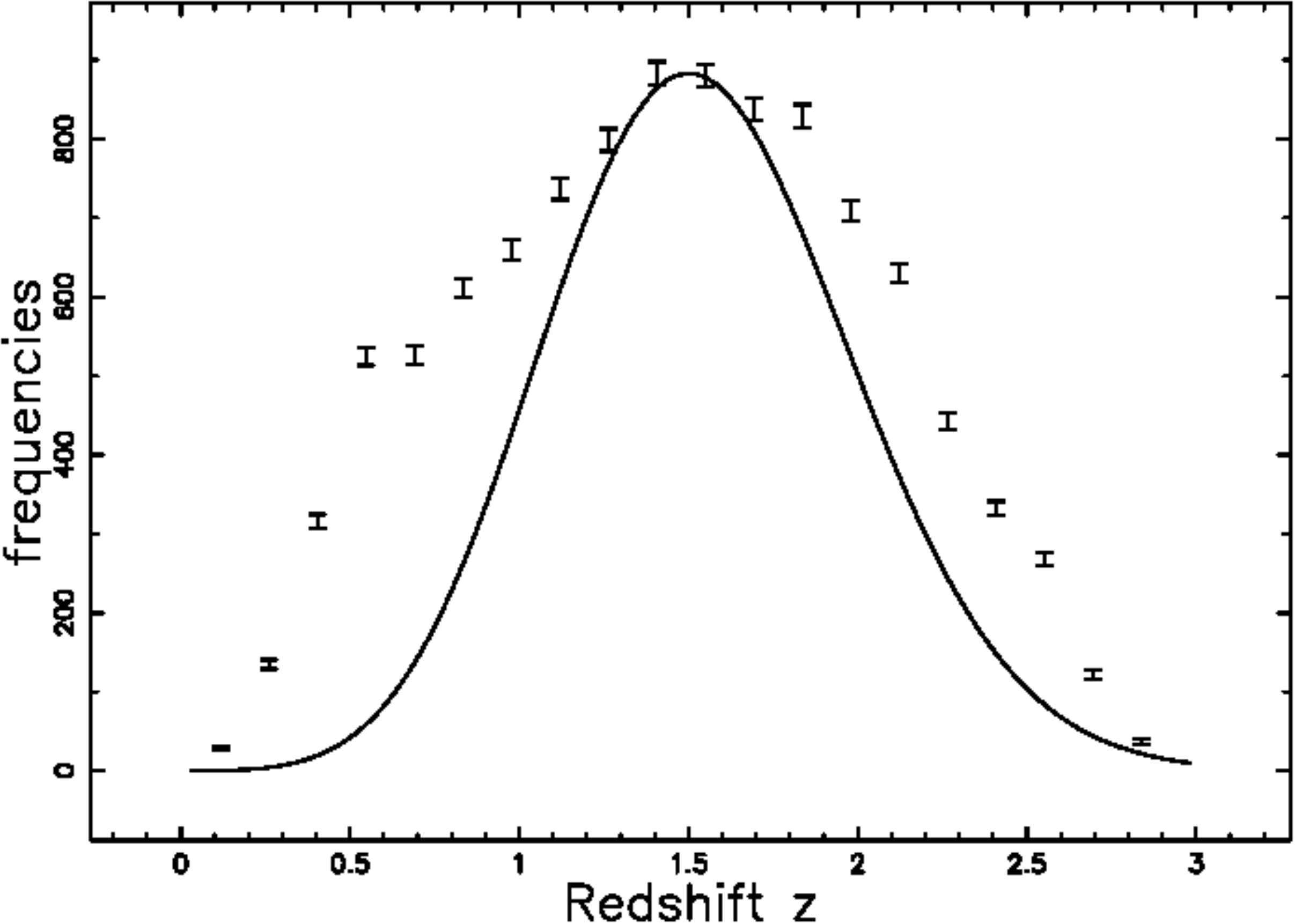}
\end {center}
\caption{
The QSOs    with
$ 20.16  \leq  m   \leq 20.85 $  
are  organized in frequencies versus
spectroscopic   redshift, points with error bar.
The redshifts cover the range $[0,3]$ and  the
histogram's interval  is 0.14.
The maximum frequency of observed QSOs is
at  $z=1.478$ 
and the number of bins is 20.
The full line is the theoretical curve
generated by
$\frac{dN}{d\Omega dz dm}(z)$
as given by the application of the truncated Schechter LF
which  is Eq.~(\ref{nzapparent}) 
with parameters as in Table \ref{trschechterfit}
but $M^*=-22.5$.
The theoretical maximum is at $z=1.491$.
}
          \label{maximum_qso}%
    \end{figure}

The number of galaxies in $z$  and $m$ as given by
formula~(\ref{nzapparent})
has a maximum  at  $z=z_{pos-max}$
but there is no analytical solution for such a position
and a  numerical analysis should be performed.
Figure \ref{maximum_qso} reports the observed and the 
theoretical number of QSOs 
as functions of  the redshift 
at a given apparent magnitude
when $M_l(z)$  is given by eqn~(\ref{mlz})
and  $M_u(z)$  is given by eqn~(\ref{absolutemagfzupper}).
Here we adopted the law of rare events, i.e. the Poisson distribution,
in which the variance is equal to the mean, i.e.
the error bar is given by the square root of the frequency.

%begin figure maximum_qso
%end  figure maximum_qso
In  the above fit the observed position of the maximum
, $z=1.478$, and the theoretical prediction 
, $z=1.491$, have approximately the same value.
In the two regions surrounding the maximum,
the degree of prediction is not as accurate,
due to the fact that
the three absolute magnitudes $M_l$, $M_u$ and  $M^*$ 
are functions of $z$.

\section{Conclusions}

{\bf Absolute Magnitude }

The evaluation of the absolute magnitude of a QSO
is connected  with  the distance modulus,
which,  in the case of 
the flat  cosmology,
($\om=0.3$, $H_0=70 \h0units$) 
is reported in eqn~(\ref{modulusflat})
as  
a Taylor series of order 8
with range in $z$, $[0-4]$.
As an application of the above series, we derived 
an inverse formula for the redshift as a function of 
the luminosity distance and an approximate 
formula for the total comoving volume.

{\bf Truncated Schechter LF}

The Schechter LF is characterized  by three parameters:
$\Phi^*$, $\alpha$  and $M^*$.
The truncated Schechter LF is characterized  by five  
parameters:
$\Psi^*$, $\alpha$, $M^*$,  $M_l$ and $M_u$.
The reference LF for QSOs, the double  power law LF,
is characterized  by four parameters:
$\phi^*$, $\alpha$, $\beta$ and  $M^*$.
An application of the above LFs 
in the range of z   $[0.3,0.5]$
gives  the following   reduced chi-square
$\chi_{red}^2=$ 2.57   for the truncated Schechter LF,
$\chi_{red}^2=$ 1.49   for the Schechter    LF,
$\chi_{red}^2=$ 1.57   for the double power LF,
and 
$\chi_{red}^2=$ 2.05   for the Pei LF.
The other statistical such as the AIC are reported
in Tables 
\ref{trschechterfit}, 
\ref{schechterfit},
\ref{doublepowerfit},
and \ref{pei14fit}. 
We can therefore speak of minimum differences 
between the four LFs here analyzed 
in the nearby universe 
defined by redshifts  $[0.3,0.5]$.

{\bf Evolutionary effects}

The evolution of the LF for QSOs  as a function of the redshift
is here modeled by an upper and lower truncated Schechter
function.
This choice allows modeling the lower bound in luminosity
(the higher bound in absolute magnitude) 
according to the
evolution of the absolute magnitude, see 
Eq.~(\ref{absolutemagfzupper}).
The evaluation of the upper  bound in luminosity
(the lower  bound in absolute magnitude)
is  empirical 
and is reported in eqn~(\ref{mlz}).
A variable value of $M^*$ with $z$ in the case 
of the truncated Schechter LF, see  
eqn~(\ref{mstarcorrection}), allows matching the evolution 
of the observed  average  value of absolute magnitude 
with 
the theoretical   average  value of absolute magnitude,
see Figure \ref{qsoxmz}.
A comparison is done with 
the theoretical average  value in absolute magnitude 
for  the case of a double power law and the Pei function,
see Figures  \ref{qsoxmz_double} and \ref{qsoxmz_pei14}.

{\bf Maximum in magnitude}

The joint distribution in redshift and energy flux density 
is here modeled in the case of a flat  universe,
see formula \ref{nfunctionzsscztrunc}.
The position in redshift of the maximum in the number
of galaxies 
for a given flux or apparent magnitude 
does not have an analytical expression
and is therefore found numerically, 
see Figure \ref{maximum_qso}.
A comparison can  be done  
with the number of galaxies
as a function of the redshift  in 
$[0-0.3]$
for the 
2dF Galaxy Redshift Survey in the South and North galactic poles, 
see Figure 6 in \cite{Cole2005}
where  the theoretical model is obtained
by the generation of  random catalogs.

%\bibliography{biblio}

\end{document}